\documentclass[a4paper,fleqn,usenatbib]{mnras}

\usepackage[T1]{fontenc}
\usepackage{ae,aecompl}
\usepackage{graphicx}	
 \usepackage{longtable}
 \usepackage{tabularx} 

\usepackage{savesym}
\usepackage{amsmath}
\savesymbol{iint}
\savesymbol{iiint}
\usepackage{txfonts}
\restoresymbol{TXF}{iint}
\restoresymbol{TXF}{iiint}

\title[Precise and powerful chaos of the 5:2 resonance]{The Precise and Powerful Chaos of the 5:2 Mean Motion Resonance with Jupiter}

\author[N. Todorovi\' c]{
Nata\v sa Todorovi\' c \thanks{E-mail:ntodorovic@aob.rs}
\\
Belgrade Astronomical Observatory, Volgina 7, P.O.Box 74 11060 Belgrade, Serbia
}

\date{Accepted XXX. Received YYY; in original form ZZZ}

\pubyear{2016}

\begin{document}
\label{firstpage}
\pagerange{\pageref{firstpage}--\pageref{lastpage}}
\maketitle

\begin{abstract}
 This work reexamines the dynamics of the 5:2 mean motion resonance with Jupiter located in the Outer  Belt  at $a\sim 2.82$ AU.
First, we compute dynamical maps revealing the precise structure of chaos inside the resonance. 
Being interested to verify the chaotic structures  as sources of natural transportation routes,
we additionally integrate 1000 massless particles initially placed along them
and follow their orbital histories up to 5 Myr. 
As many as 99.5\% of our test particles became Near-Earth Objects, 23.4\% migrated to 
semi-major axis below 1 AU and more than 57\% entered the Hill sphere of Earth. We have also observed a borderline 
 defined by the  $q \simeq 2.6$ AU perihelion distance along  which particles escape from the Solar System. 
\end{abstract}




\section{Introduction}

Some of the most important resonances in our Solar System
are  those capable of  driving bodies from different parts of the Main Belt down to the neighborhood of Earth, to
the so called Near Earth Object (NEO) region. By convention, NEO region is defined for perihelion distances smaller than $q\leq 1.3$ AU
and aphelion distances larger than $Q\geq 0.983$ AU \citep{Rabinowitz1994}.

A generally accepted dynamical scenario played by those 'important' resonances evolves according to the following principle:
the asteroid is slowly driven into the resonance  by the Yarkovsky effect \citep{VokrFarin2000} or becomes  directly
injected into  it by some collisional event. The semi-major axis of the resonant asteroid stays almost 
constant, while its eccentricity slowly increases up to planet crossing values \citep{Wetherill1979, Wisdom1983}, opening
possibilities for close encounters. Depending on  its proximity to the planet and  mass, 
the asteroid may or may not survive the close encounter. 
We do not go into detailed description of all the possible close encounter outcomes, but in general, 
the small body continues its journey through  the phase-space governed by the subsequent close encounters or captures into
other resonance/s, until it collides with a planet or the Sun, becomes ejected to hyperbolic orbits or drives out of the Solar system. 

First systematic numerical studies on the most prominent Main Belt resonances performed by \cite{GladmanSCIENCE1997},
showed that the evolutionary processes driving bodies to the NEO region  unfold mostly 
due to three Main Belt resonances:
the $\nu_6$ secular resonance located at $a \sim 2.1$ AU, 3:1  and 5:2 mean motion resonances (MMRs) 
with Jupiter, at $a \sim 2.50$ AU and  $a \sim 2.82$ AU respectively. 
The 5:2 resonance, the subject of our study, was considered as the least efficient in that  respect,
since most particles placed into that  resonance, after approaching Jupiter,
were ejected out to hyperbolic orbits.
Similar results  were obtained  in another round of numerical studies  by  \citet{Bottke.et.al.2002}, 
who claimed that  only 8\% of the NEO population comes from the outer belt 
($a >2.8$ AU which includes the region of the 5:2 resonance). Furthermore, \citet{EliaBrunini2007} 
estimated that as many as 94\% of NEAs (Near Earth Asteroids)  arrived from the Main Belt region inside the 
5:2 MMR, confirming that the Outer Belt has a negligible role in the dynamical processes supplying the NEO region.

On the other hand, spectral analysis suggests that a large number of  ordinary L chondrite meteorites     
could have been driven down to Earth from the outer belt in very short delivery times (<1-2 Myr).
The main candidate for the asteroidal source of L-chondrite meteorites is the  
outer belt asteroid family Gefion, whose fragments, after a catastrophic breakup $\sim 470$ Myr ago, rapidly evolved 
to Earth-crossing orbits via the nearby 5:2 mean motion resonance with Jupiter \citep{Nesvorny2009}. 
However, the authors asserted that 'a potential problem with this result 
should be the low efficiency of meteorite delivery from the Gefion family location'.

Asteroid (3200) Phaethon,  the parent body of the Geminids meteor shower, arrived in the NEO region
most likely from the asteroid family Pallas (located in the Outer Belt), via its bordering 8:3 or 5:2 MMRs with Jupiter \citep{deLeon2010}.
Still, dynamical models in \cite{deLeon2010, Bottke.et.al.2002} showed that less than $\sim 1\%$  of the test particles placed into the
5:2 resonance  recovered a Phaethon-like orbit.  
We mention another candidate that most likely arrived from the Outer Belt Koronis family via the 5:2 resonance -
one of the largest Potentially Hazardous Asteroids - 2007 PA8 \citep{Sanchez2015ApJ, Nedelcu2014}.

Thus, a growing number of meteorites and NEOs whose spectral type indicates
an Outer Belt origin  and which have been driven to the NEO region via the 5:2 resonance,
does not match well with its low transportation efficiency suggested in dynamical studies.

Let us notice that a common feature in the studies of \cite{Bottke.et.al.2002, GladmanSCIENCE1997, MorbGlad1998,
EliaBrunini2007} and \cite{deLeon2010} 
is that  unbiased initial data sets were used, with no particular selection  from the most unstable parts of the resonance. 
Moreover, \cite{GladmanSCIENCE1997} claimed that the initial location inside the resonance should not have a significant influence on the 
orbital scenaria.

Here we reexamine the transportation abilities of the 5:2 resonance
using the same principle as in the above mentioned studies, with the exception that the test particles are carefully chosen at the 
most unstable parts of the resonance, whose localization is enabled by the computation of highly precise FLI (Fast Lyapunov Indicator) dynamical maps.

The article is organized as follows.
In Section \ref{FLI} we give a brief definition of the Fast Lyapunov Indicator and in Section \ref{FLImaps} we show and describe the FLI
maps we have computed using a Solar System model  with and without the inner planets.  A short description of the way we select the 1000 particles
in order to confirm the high transportation abilities of the 5:2 resonance is given in Section \ref{method}. 
The resulting orbits  are discussed in Section \ref{results}, while the Conclusions are provided in  Section \ref{concl}. 
In the supplementary data (on-line only) we list the exact initial values of the orbital elements of  1000
rapidly evolving particles and show some additional results on the 5:2 MMR.

\section{The Fast Lyapunov Indicator-FLI}
\label{FLI}
The Fast Lyapunov Indicator \citep{FLGonz1997, FGonzL1997} is one of the most efficient chaos detection tools
used mainly to produce dynamical maps. In the beginning FLI was used for idealized systems such as symplectic maps or simplified Hamiltonians 
\citep{Froeschle2000Science,LGFArndiff2003, FGLLocGlobdiff2005, Todo2008, Todo2011}, but later FLI was successfully applied in studies 
of asteroids or planetary systems as well,
starting from the close neighborhood of Earth \citep{Daquin2016, Rosengren2015}, up to the  outer Solar System  \citep{Guzzo2005, Guzzo2006}. 
FLI maps were also produced for the region  between Earth and Venus by \cite{Bazso2010},  Main Belt 
maps were computed in \cite{Gales2012}, investigation of the asteroid family Pallas  based on FLI maps was 
performed in \cite{TodoNov2015} and finally, exoworlds have also been extensively studied by FLI, for example in
\cite{PilatDvorak2002, Dvorak2003, Sandor.etall2007, Schwarz2011}.  What follows is a brief definition of FLI.

Let us consider a continuous dynamical system defined by a set of differential equations
\begin{equation}
 \frac{dx}{dt}=\mathcal{F}(x)
 \label{eq:motion}
\end{equation}
where  $\mathcal{F}: M \rightarrow  M$ is  a differentiable function defined on  a  manifold $M \in \mathbf{R}^n.$
For a given initial value $x(0)$ and its corresponding initial nonzero  deviation vector $v(0)$  lying in the 
tangent space $\mathcal{T}_x M$  of $M$,
the  Fast Lyapunov Indicator is defined 
as the logarithm of the  norm of the deviation vector $v(t)$ at some fixed time $T$, i.e. by the quantity:
\begin{equation}
FLI (x(0),T) = \sup_{t \leq T} \log \| v(t) \|
\label{FLIdef}
\end{equation}

In the above definition time $T$ plays the role of a parameter, while the supremum is used only to annul some local
oscillations that may influence the result.

When it comes to the numerical evaluation of FLI, one has to follow both the orbit $x$ and its corresponding deviation vector $v$.
Time evolution of $x$ is obtained by integrating the
 equations of motion (\ref{eq:motion}), while the evolution of $v$ is released by integrating the so-called variational equations:
$$\frac{dv}{dt}=\frac{d\mathcal{F}}{dx}(x(t)) v, $$
where $d_x\mathcal{F} ^t$ is an operator  which maps the tangent space of $M$
at point $x$ onto the tangent space $\mathcal{T}_{\mathcal{F}^t(x)} M$ at point $\mathcal{F}^t(x)$.

If the orbit $x$ is  regular,  the norm of its deviation vector  $\| v(t) \|$  grows  linearly  over time. 
For  chaotic orbits $\| v(t) \|$ grows  exponentially. Stronger chaos implies its
faster increment and hence larger FLI values.

Using FLI basically means that we do not have to wait a long time  for the orbit to show its dynamical character through
the asymptotic properties of its deviation vectors. The integration time should be short enough just to capture the difference between resonant, 
nonresonant or chaotic orbits of  different strengths. 

Moreover, \cite{GuzzoLega2014}  illustrated that tangent vectors grow faster  
if the  initial condition of the orbit is close to stable-unstable manifolds of the hyperbolic fixed point 
of a resonance. Such orbits will have small FLI ridges that can be captured only at the beginning of the integration. 
Therefore, calculation of FLI maps with  a careful choice of the computing time allows a clear visualization of the 
hyperbolic structures inside the resonance, opening up many possibilities for their further investigation.

\section{FLI maps of the $5:2$ resonance with Jupiter}
\label{FLImaps}

\begin{figure}

 \hskip -2.cm \includegraphics[height=12.7cm, angle=-90]{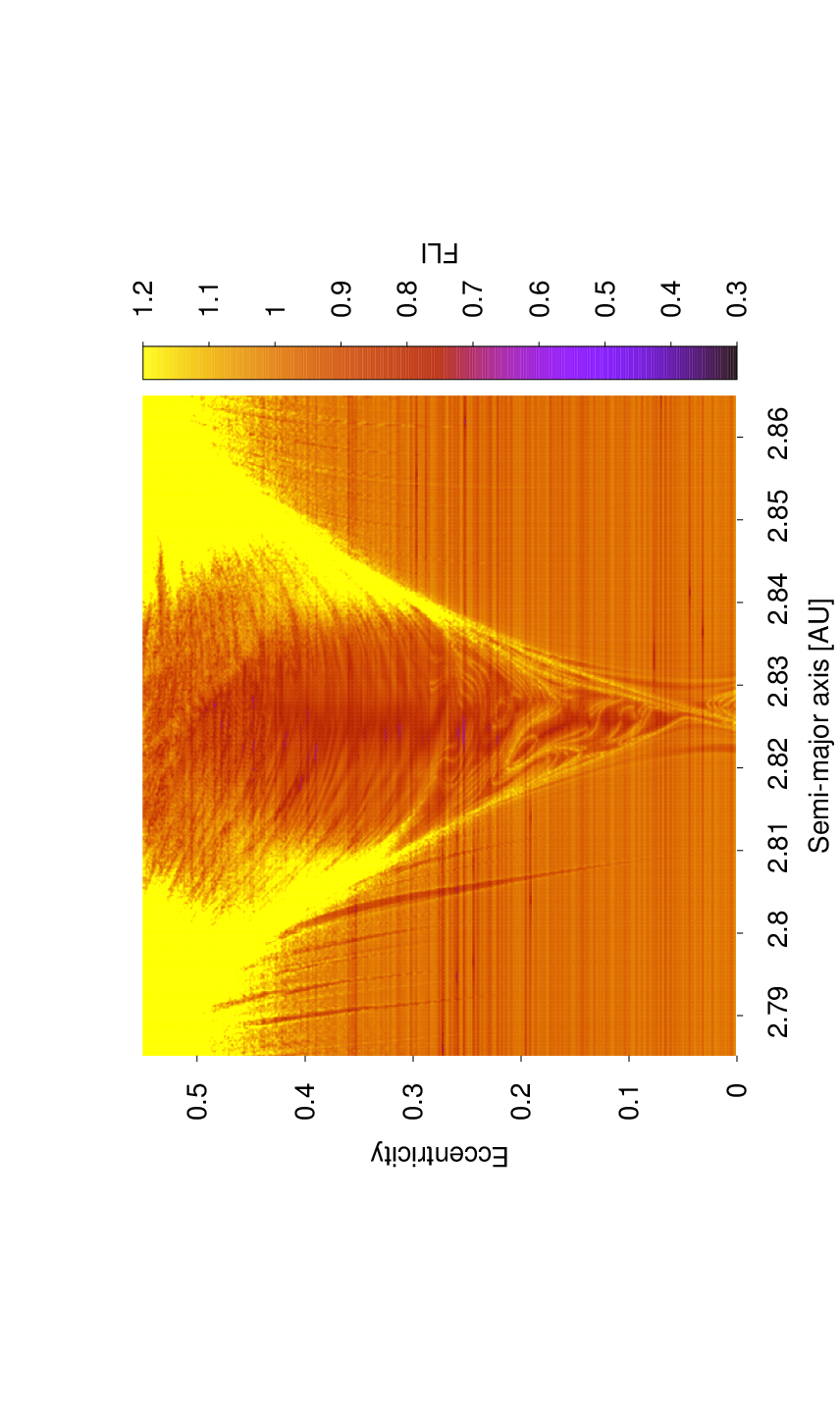}\\

 \hskip -2.cm \includegraphics[height=12.7cm, angle=-90]{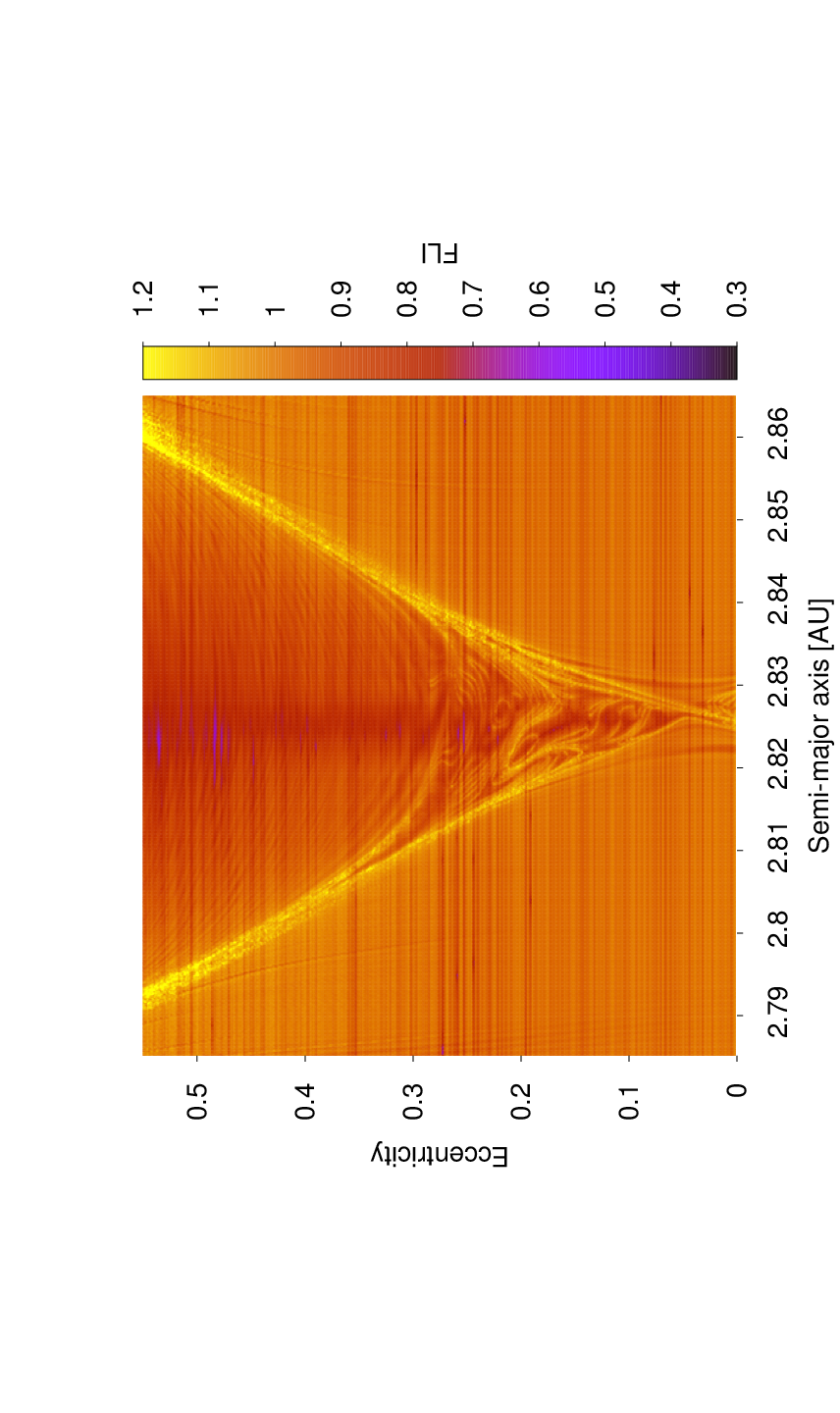}
\caption{The region of the 5:2 mean motion resonance with Jupiter computed with FLI using the
full Solar System model (top) and outer planets only (bottom).
In the upper figure, chaos dominates in the region at larger $e$. 
The separatrix becomes misshapen at $e\sim 0.3$ and the region out of the resonance is crowded by weak MMRs 
whose overlapping causes most of the chaos visible in the figure.
Removing the inner planets from the model (bottom plot), cleared away the large number of MMRs
and all the chaos they generate, leading to the conclusion that MMRs visible in the top plot are 
resonances with Venus, Earth or Mars. 
Considering that the masses of the inner planets are relatively small, the amount of chaos they produce for only 10 000 years is quite large. 
The  very fine structure of chaos  inside the separatrix at lower $e$ is not affected by the presence of inner planets.}
\label{figure1}
\end{figure}

The FLI maps of the dynamical structure in the region of the  5:2  mean motion resonance with Jupiter is represented on the two panels
composing Fig.~\ref{figure1}.
The first, above panel, is calculated   taking into account  $all$ the planets in the 
Solar System, from Venus to Neptune, while Mercury's mass is added to the mass of the Sun and the  corresponding barycentric correction is applied 
to the initial conditions. The lower, bottom map is computed  including only the outer planets: Jupiter, Saturn, Uran and Neptune. 

For each  of the particles regularly distributed along  a $[500\times500]$ grid for $(a,e)=[2.785,2.865]\times [0.0, 0.55]$ 
we have computed  its corresponding FLI for 10 000 years.
Inclinations for all the particles are set to $i=10\degr$ (corresponds to the orbital plane
of the most massive asteroid Ceres), while the orbital angles are fixed  at random values  
$(\Omega, \omega, M)=(260\degr, 70\degr, 99\degr)$. 
 On the color scale, stable particles with FLIs below 0.9 are red, while the most chaotic ones with $FLI> 1.1$ are yellow.
 
All the calculations are  made using the ORBIT9 integrator\footnote{Available from http://adams.dm.unipi.it/orbfit/}
that  operates on  a symplectic single step method (implicit Runge-Kutta-Gauss) as a starter and a multi-step predictor which performs most of the  propagation. 
The choice of the integrator is based on the fact that it allows not only to
integrate differential equations of motion  but also  a simultaneous integration of the corresponding variational equations, 
which are both involved in the estimation of FLI. The integrations were performed on a cluster, where the calculation time for one map is around
30 minutes\footnote{The Fermi cluster located 
at the Astronomical Observatory of Belgrade consists of 12 worker nodes, each node (HP SL390S blade server G7 X5675) has a 2xX5675(2x6core) processor,
3.1GHz, 24G memory, 2Tb disc and 2xM2090 NVIDIA Tesla-fermi GPU cards.}. 

The most evident difference between the two plots is the amount of chaos. 
 In the top panel, almost all  particles  out of the 5:2 MMR at higher $e$ are chaotic. The line of the resonant border
becomes misshapen already at $e \sim 0.3$ and a bunch of weak mean motion resonances arises from the chaotic upper part of the picture. 
In fact, a large number of resonances and their mutual overlapping generates all the chaos visible on the map.
The bottom plot lacks chaos and a dense web of MMRs visible in the top panel. However, some weak thin vertical resonant-like 
shapes are visible in the lower panel as well. We do not identify all those resonances (it is beyond the scope of this paper and can be the 
subject of some future research)
but a direct comparison between the two plots leads to the conclusion that resonances visible only in the 
top figure  are MMRs generated by the inner planets - Venus, Earth or Mars. Following the same logic,
weak resonances visible only in the bottom plot are caused by some of the outer planets.
 
We certainly do expect that the full Solar System model generates more chaos than the incomplete one,
but having in mind the relatively small masses of the inner planets, 
the amount of chaos they produce for only 10 000 years is quite large.

The structure of chaos  is clearly detected for eccentricities bellow $e \sim 0.3$,
where many peculiarly shaped thin yellow lines are noticeable.
Since their position and form  remains unchanged on both plots, we conclude that 
the fine structure of chaos at low eccentricities is not affected by the presence of inner planets.

We  consider a realistic model, which means that the observed structures should be placed exactly on the location we see on the plot. 
On the other hand, the orbital space is 6-dimensional, while the maps are 2-dimensional, 
i.e. the plots give a realistic, but only a partial insight into the chaotic structure of the 5:2 MMR.
We notice that the peculiar structures  look very similar to the traces of the 
hyperbolic invariant manifolds of the saddle point of the resonance, observed for example in  \citet{GuzzoLega2014, GuzzoLega2015}. 
However, potential affiliation of those structures to the hyperbolic set requires a more detailed study and stronger theoretical grounds. 

\section{Evolution of 1000 test particles}
\label{method}

The principal  aim of this research is not only to illustrate the beauty of chaos in one 
of the most important Main Belt resonances, but also to show that the  most unstable parts of the resonance
do provide a fertile source of rapid delivery routes. For that purpose, we have chosen 1000 test particles 
initially placed along the  chaotic structures  visible on Fig.~\ref{figure1}
and integrated them up to 5 Myr. We chose particles  knowing a priori
they would be 'active' as soon the integration starts. In this way a significant reduction of calculation time can be achieved, 
since nominal integration times used in similar studies \citep{GladmanSCIENCE1997, Bottke.et.al.2002, deLeon2010}
were much longer, up to 100 Myr. 
Initial  semi-major axis and eccentricities of the particles are between $a\in [2.820, 2.834]$ and $e \in [0, 0.22]$.
Initial values of the remaining orbital elements are the same as for the computation of Fig.~\ref{figure1}, that is
$(i, \Omega, \omega, M)=(10\degr, 260\degr, 70\degr, 99\degr)$. 

The integrations were stopped if the orbit became  hyperbolic or the particle  reached semi-major axis larger than $a >100$ AU.
The software used for integration (ORBIT9) deals well with planet close encounters, but it can not simulate collisions. Therefore,
eventual planet-particle or particle-particle collision end states are not registered in the integration. 

 It should be noted that  all our test bodies lie in the same inclination plane and have the same orbital angles. Therefore,
the results presented reflect the migration abilities of a very small portion of the resonance.
Selecting  particles  along all the 6 dimensions (or at least for different sets of the remaining 4 variables)
would provide a more generic result on the 5:2 MMR diffusion capacities, but this approach is computationally very demanding, 
goes beyond the scope of this paper and  will be the subject of further investigation.

\section{Results}
\label{results}
Considering  that the 5:2 MMR is one of the best studied resonances in the Solar System, we will not repeat the known results on its dynamics, 
but rather focus on some new aspects of its transportation abilities. Since some results show a very 
large disagreement with previous studies on the 5:2 MMR, in addition (Table \ref{table1}  in supplementary data) we list the exact values 
of the initial orbital elements of the 1000 test particles, so that the reader can reproduce the same integration. One part of Table \ref{table1}
is given below. 

\begin{table}
\centering\begin{tabular}{c c c c}
\\
\hline\hline 
 $a_0$    &  $e_0$  &  $a_0$    &  $e_0$    \\
\hline
\hline
2.8270400   & 0.0504400     &  2.8329999   & 0.1742000    \\
2.8270800   & 0.0504400     &  2.8339601   & 0.1960400    \\
2.8271201   & 0.0410800     &  2.8330400   & 0.1726400    \\
2.8272400   & 0.0400400     &  2.8330801   & 0.1773200    \\
2.8273201   & 0.0416000     &  2.8330801   & 0.1976000    \\
2.8273602   & 0.0431600     &  2.8331201   & 0.1747200    \\
2.8273602   & 0.0436800     &  2.8331201   & 0.1762800    \\
2.8274400   & 0.0452400     &  2.8331201   & 0.1768000    \\
2.8274400   & 0.0457600     &  2.8331201   & 0.2033200    \\
2.8274801   & 0.0520000     &  2.8205600   & 0.1424800    \\
2.8259602   & 0.0416000     &  2.8332000   & 0.1768000    \\
2.8259602   & 0.0514800     &  2.8332000   & 0.1773200    \\
2.8251200   & 0.0078000     &  2.8332000   & 0.1996800    \\
2.8260000   & 0.0421200     &  2.8332400   & 0.2043600    \\
2.8260000   & 0.0452400     &  2.8332801   & 0.2142400    \\
2.8205600   & 0.1430000     &  2.8268800   & 0.0171600    \\
2.8206401   & 0.1414400     &  2.8269200   & 0.0078000    \\
2.8209600   & 0.1341600     &  2.8275201   & 0.0114400    \\
2.8210402   & 0.1326000     &  2.8262801   & 0.0192400    \\
2.8208802   & 0.1357200     &  2.8226800   & 0.1492400    \\
2.8218000   & 0.1253200     &  2.8267601   & 0.0161200    \\
2.8205600   & 0.1424800     &  2.8262401   & 0.0052000    \\
\hline
\end{tabular}
\caption{Initial semi-major axis and eccentricities of some test particles used in our integration.
The complete  Table 1 (containing initial values for all 1000 particles) is given in supplementary material. 
The initial inclination, longitude of the node, longitude of the perihelion  and the mean anomaly are fixed on 
$(i, \Omega, \omega, M)=(10\degr , 260\degr, 70\degr, 99\degr)$. All the coordinates are given in the osculating orbital 
elements for the epoch JD 2456200.}
\label{table1}
\end{table}    

\subsection{Transportation to the NEO Region}

\begin{figure}
 \centering\includegraphics[height=8.cm, angle=-90]{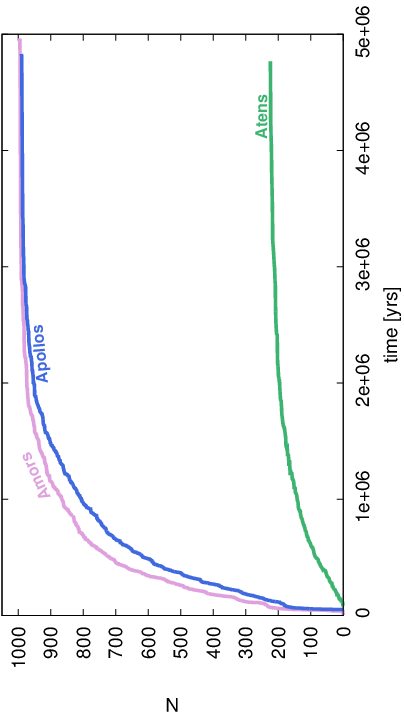}
\caption{Number of  particles (N) reaching  the region of Amors (upper curve) Apollos (middle curve) and Athens (lower curve).
First entries are recorded after 35 000, 49 500 and 82 000 years, median times of  those entries
are $0.26$ Myr, $ 0.35$ Myr and $0.68$ Myr, and the total number of bodies reaching into
the three NEO groups are 995, 991 and 225, respectively.}
\label{figure2}
\end{figure}

 As many as $99.5 \%$ of the test  bodies became NEOs. More precisely, 995  out of 1000 particles at some moment reached 
perihelion distances smaller than  $q<1.3$ AU.  Statistical representation of the entry rates into the three NEO 
groups: Amors ($1.0167 AU \leq q \leq 1.3 AU$), Apollos ($a \geq 1.0 AU, q \leq 1.0167 AU$) 
and Athens ($a < 1.0 AU, Q \geq 0.983 AU$) is  given on Fig. \ref{figure2}.

The large amount of material migrating from the 5:2 resonance into the NEO region was not observed in earlier studies.
For example, in \cite{GladmanSCIENCE1997, Bottke.et.al.2002} or \citet{EliaBrunini2007} this percentage was less than 10\%.
Such disagreement is primarily attributed to the way we  choose initial conditions. 

\begin{figure}
 \includegraphics[height=8.5cm, width=9cm, angle=-90]{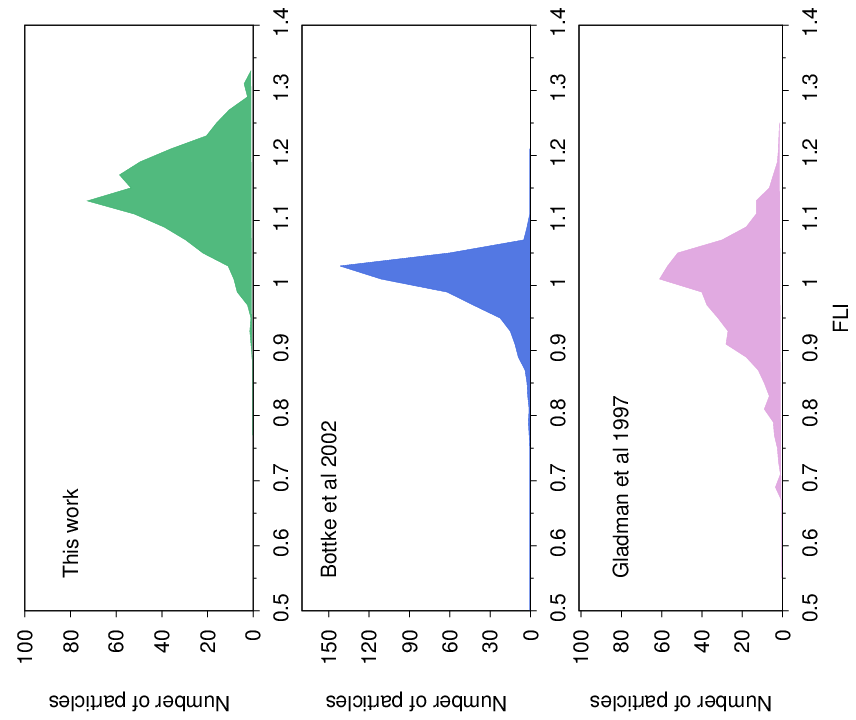}
\caption{Distribution of FLI values of the particles used in this work (top), in Bottke et al 2002 (middle) and  Gladman et al 1997 (bottom).
The FLIs on the top plot have systematically larger values than the ones from the two plots below. 
Accordingly, they should have stronger migration abilities. }
\label{figure3}
\end{figure}

In order to illustrate this, we compare the chaoticity of test particles used in different studies. 
That is, we compute the FLI values of the initial data set used in this work, 
with the FLIs of the  particles that were chosen in the same way as in \citet{GladmanSCIENCE1997} and \cite{Bottke.et.al.2002}. 
The three resulting histograms are given on Fig.\ref{figure3}.

Considering the work of \citet{GladmanSCIENCE1997}, we sampled 1000 test bodies in between the borders of the 5:2 resonance in the region of the 
Gefion asteroid family and computed their FLI values for 10 000 yrs. The corresponding histogram is given in the bottom panel of  Fig.\ref{figure3}.
The FLIs range between $0.55< FLI< 1.25$. Among them,  as many as  $91.5\% $  bodies have $FLI<1.1$. 
We know from Fig.\ref{figure1} that such particles are not very chaotic, i.e. they are not the best candidates to 
drift rapidly through phase-space.

In \cite{Bottke.et.al.2002} the particles were sampled over a large part of the Outer Belt. 
For a detailed description of how the test bodies  were selected, see section 2.5 in \cite{Bottke.et.al.2002}. 
Following this description, we took 1000 asteroids with $2.83< a < 2.95$ and $2.40< q <2.60$ 
from the Ted Bowell database (available at  http://asteroid.lowell.edu) and
computed their FLIs for the same amount of time, 10 000 years. The  relevant histogram is given in the middle panel of Fig.\ref{figure3}.
Here FLIs range between $ 0.25 < FLI < 1.20$ and they have somewhat smaller diversity. A strong peak appears at $FLI\sim 1.03$, 
but only 5 asteroids are strongly chaotic having $FLI > 1.1$. 

The FLI  distribution of  the 1000  particles used in this study is given on the top panel of Fig.~\ref{figure3}.
FLIs range between $ 0.77 < FLI< 1.33$, but as many as $75.2\%$ bodies have FLIs larger than $ FLI> 1.1$, i.e. a large majority is very chaotic
justifying their fast migration abilities.

Time scales at which the 5:2 MMR increases eccentricities up to planet crossing values are similar to the ones observed in earlier studies. 
In both semi-analytical  \citep{Yoshikawa1991}  and numerical studies \citep{Ipatov1992, MorbGlad1998} those times where estimated on $\sim 10^5$ years. 
Here, median time, i.e. time at which 50\% of the particles became NEOs, is $T_{med}=2.5\cdot10^5$ yrs.  The 
mean time (arithmetical mean) of the first entry into the NEO region is $T_{mean}= 4\cdot10^5$ yrs.

\begin{figure}
\centering\includegraphics[height=8.cm, angle=-90]{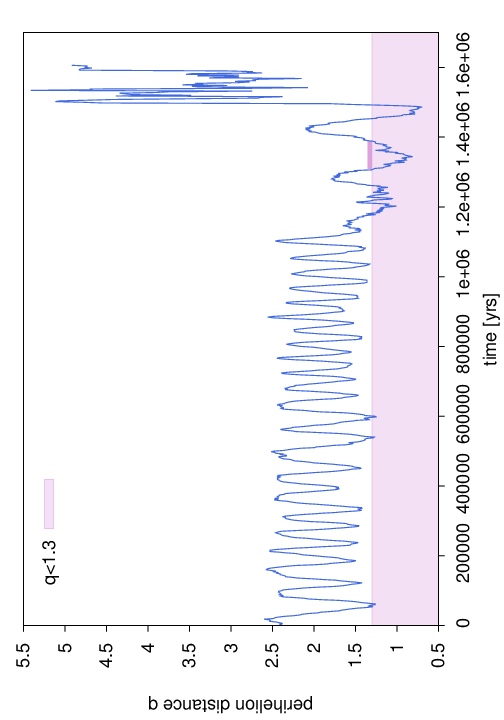}
\caption{Random walk of the perihelion distance for one particle illustrating  multiple entries into the NEO
region (shadowed area). The residence time after its first entry (at $\sim 50 000$ years)  is only a few 100 years. 
The longest stay below the $q=1.3$ AU line is 80 000 years and is marked with a bold line starting at $t=1.3\cdot10^6$ yrs.}
\label{figure4}
\end{figure}

 We are also interested to estimate the mean lifetime the objects spend in the NEO region. 
This estimation is often uncertain because most particles make multiple entries (and reentries).
The situation is illustrated in Fig. \ref{figure4}, where we can follow  the irregular behavior of the  perihelion distance $q$ for one particle,
along with its numerous entries into the NEO region (marked by the shaded area). 
The first time the object became a NEO was at $t\sim 50 000$ yrs, but only for a few 100 yrs.
The longest time it persisted  below the NEO border was 80 000 yrs. In Fig. \ref{figure4}, this interval is marked with a bold line (at $t\sim 1.3$ Myrs). 
The mean value of the {\it first} stay in the NEO region is relatively short, $T_{mean}= 37 000$ yrs. The mean value of the {\it longest} stay 
is $T_{mean}=0.2$ Myr,  comparable with the result from \cite{Bottke.et.al.2002} where this time was estimated on $T_{mean}= 0.19$ Myr. 

\begin{figure}
\centering\includegraphics[height=8.cm, angle=-90]{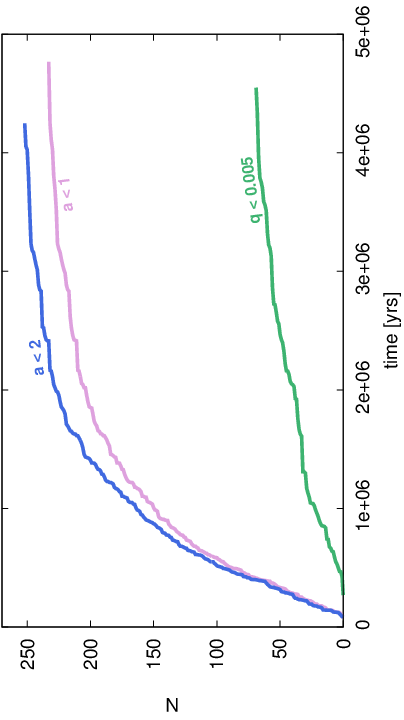}
\caption{Number of particles (N) migrating down to smaller semi-major axis: to $a<2$ AU (top curve),
to $a<1$ AU (middle curve), and to perihelion distances with Sun grazing values $q<0.005$ AU (bottom curve).  
The first entries into the three regions are registered after 80500, 81000 and 251000 years, 
median times of the entries are $0.67$ Myr, $0.66$ Myr and  $1.63$ Myr,
and the total number  of particles reaching the three criteria are 253, 234 and 70, respectively.}
\label{figure5}
\end{figure}

\subsection{Migration to the Sun}

A large number of asteroids migrated closer to the Sun as well. 
Here we are able to directly compare (Table \ref{table2})  our results with the ones 
from  \cite{GladmanSCIENCE1997} and \cite{Bottke.et.al.2002}.
According to those studies, the 5:2 MMR was not capable of driving bodies to semi-major axis below $a<1$ AU.  
In our integration, this result is significantly different,  as many as $23.4\%$ of the particles entered  the $a<1$ AU region. 
A similar situation is observed for the orbits with $a<2$ AU, $25.3\%$ of the bodies at some point reached semi-major axis below $2$ AU,
while in \cite{Bottke.et.al.2002} this fraction did not exceed $6\%$. 

\begin{table}
\begin{tabular}{ l  c  c c}
\hline\hline
                                & \shortstack{This \\ work}  &  \shortstack{Gladman\\ et al 1997} & \shortstack{Bottke \\ et al 2002} \\
           \hline
 Number of particles            &     1000       &        146                 &      359        \\
 Integration time (Myr)         &        5       &        100                 &      100         \\
 Ever have $ a<2 $ AU  (\%)     &     25.3       &        1.4                 &       <6          \\
 Ever have $ a<1$  AU  (\%)     &     23.4       &        0                   &       -          \\
 Ever have $ q<0.005$ AU (\%)   &      7         &        7.5                 &       -          \\           
\hline   \hline
\label{table2}
\end{tabular}
\caption{The fraction of particles reaching  semi-major axis values lower than $a<1$ AU, $a<2$ AU and perihelion distances $q<0.005$ AU observed in this 
work and in  Gladman et al (1997)  and Bottke et al (2002). We also give the total number of particles and the integration times in all the three studies.}
\end{table}

The rate of Sun-grazing  bodies, i.e. bodies that have perihelion distances smaller than $q<0.005$  is $7\%$. 
Among the 70 Sun grazers, 69 got there by increasing eccentricities close to 1, while only one particle actually approached the Sun due to the decrement of the semi-major 
axis down to 0. Therefore we can conclude that low values of $q$ are not affected by the decrease of $a$. This could 
explain why we obtained almost the same result as \cite{GladmanSCIENCE1997}, where no particles with low $a$ have been observed,
but the percent of bodies with $q<0.005$ was $7.5$\%, close to the value obtained in this study. 

The entry rates into the regions with $a<1$ AU, $a<2$ AU and $q<0.005$ AU are given on Fig.\ref{figure5}.

\subsection{Close encounters}
             
 \begin{table}
\begin{tabular}{m{0.3cm}  l  m{2cm}  c    m{0.3cm} }
 \hline
 \hline                                                    
 &    Planet         &       &    (\%)      &     \\
 &                   &       &              &     \\
\hline
 &    Venus          &       &     46.0     &     \\
 &    Earth          &       &     57.4     &     \\
 &    Mars           &       &     63.0     &     \\
 &    Jupiter        &       &     98.5     &     \\
 &    Saturn         &       &     67.3     &     \\
 &    Uran           &       &     42.6     &     \\
 &    Neptune        &       &     25.1     &     \\
 &    no close enc.  &       &      0.6     &     \\
  \hline 
   \hline 
 \end{tabular}
 \caption{ The percentage of bodies entering the Hill spheres of each planet in the model. 
 Jupiter dominates with 98.5\%. The number of bodies approaching other planets decreases 
 gradually as we move away from the resonance, both inwards and outwards the Solar System, while
 6 particles avoid any close encounters. }
 \label{table3}
 \end{table}

As mentioned above, the software we use does not register collisions
(we do not treat any physical parameters required for those estimates). 
Therefore we 'simulate' high collision probabilities by decreasing close approach distances down to 0. 
 Our integrations are performed with nominal close approach distances 
($d=0.01$ AU for inner planets and $D=1$ AU for outer planets). Since an enormous number of 
close encounters was registered,  we repeat the integration for smaller values of $d$ and $D$.

That is, we set close approach distances to the radius of the  Hill spheres of the planets
($d=0.007$ AU for inner planets and  $D=0.7$ AU for outer planets) and count the number of bodies 
entering the Hill spheres. It should be noted that changing close 
approach distances does not affect any other aspect of the evolved orbits (except that it counts close encounters), and that one particle
usually has several planet rendezvous before it ends its journey through the Solar System.

In Table \ref{table3} we give the percentage of bodies entering the Hill sphere of each planet in the model.
As expected, Jupiter dominates with  $98.5\%$  bodies approaching it.
The large amount of bodies getting close to it should not be a surprise,
since most particles placed in the 5:2 MMR are ejected out of the system by Jupiter. 
The number of bodies approaching other planets decreases gradually as we move away from the resonance, both 
inwards and outwards the Solar System, while 6 particles stayed in the resonance avoiding any close encounters. 

The amount of bodies entering the Hill sphere of Earth is very high, $57.4\%$. 
This result was reconsidered in a new round of numerical integration, where the close approach distances were set to $d=0.001$ AU.
We have found 5 bodies approaching Earth that close. The integrations were repeated for smaller  values of $d$, until no close encounters
were registered. The lowest such  limit  was $d=0.0007$ and two bodies were found to have approached Earth that close.

Although we were not able to estimate exact Earth collisional probabilities, the above result suggests that,
if we choose particles at the most unstable parts of the resonance, the 5:2 resonance has at least one order of magnitude 
higher collisional probability with Earth, than the one observed in \cite{MorbGlad1998} or \cite{EliaBrunini2007}, 
where this value was estimated on $P_{col}\sim 10^{-4}$.

\subsection{Final destinations}

\begin{figure}
\centering \includegraphics[width=9.cm, height=15cm, angle=-90]{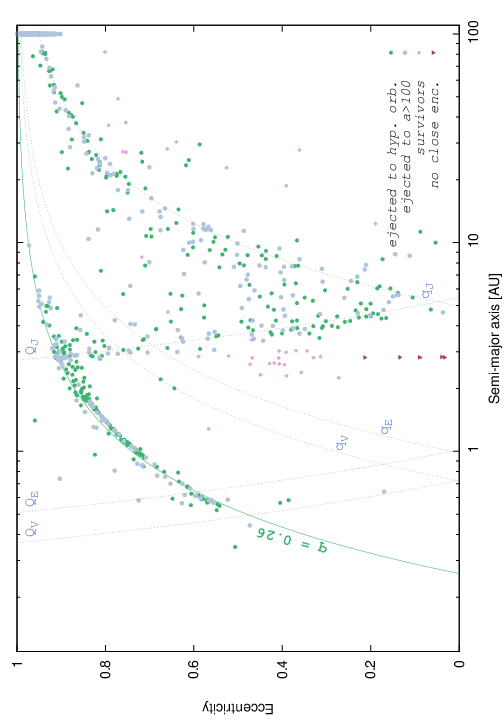}
\caption{Final distribution of the 1000 test particles in the $(a,e)$ plane.
The green and blue  dots are the positions from which the particles  were ejected to hyperbolic orbits and orbits with  $a>100$ AU.
Those removals are clearly concentrated  along and in between the perihelion ($q_J$)  and aphelion ($Q_J$)  lines of Jupiter.
An undefined source of removals is the sickle shaped line following a 
direction whose best numerical fit corresponds to $q\simeq 0.26$ AU (marked with a green full line). 
Since no secular resonances were found at this  location, the $q\simeq 0.26$ AU course may represent a natural stability 
border of the inner Solar System.
The survivors are marked with pink squares and the purple  triangles are the positions of the 5 particles that 
have remained in the resonance avoiding close encounters.}
\label{figure6}
\end{figure}

End states of the integrated particles are also under consideration. Distribution of their final positions in the $(a,e)$ 
plane is given in Fig.~\ref{figure6}. 
The pink squares are the positions of the 42 survivors at the end of the integration.
The 5 particles that stayed in the resonance and had not managed to raise up to planet crossing values are marked with purple triangles.
The  green dots (320 of them) are the locations from which the particles were ejected to hyperbolic orbits and the blue ones (638 of them) 
are the positions from which the  bodies were thrown out to $a>100$ AU.

A large majority of those eliminations were caused by Jupiter, since most of the final positions are distributed along and in between the lines
of its perihelion ($q_J$) and aphelion ($Q_J$) distances. Some eliminations can be credited to other planets as well. For example, to 
Venus, the two green dots lying on the  perihelion line  $q_V$ 
at $(a,e)\sim(2,0.64)$ and $(2.2, 0.67)$  or Earth, the blue dot on the  line $Q_E$ at $(a,e)\sim(0.58,0.72)$. 
However, those removal scenaria confirm the expected and well known dynamical scheme of the 5:2 resonance. 
Median lifetimes of the particles (0.8 Myr) are in good agreement with previous results \citep{GladmanSCIENCE1997}.

The unexpected source of elimination is the sickle shaped line marked with a full green  line. 
Since we have not found any resonance on that direction, it is uncertain if those removals 
are caused by some unknown resonance or another dynamical (or even numerical) mechanism.
We have fitted a $q$ line through this removal course, which corresponds to  the  perihelion distance $q \simeq 0.26$ AU.

As mentioned above, all our test bodies  lie in the same inclination  plane and have the same values of the orbital angles.
This could mean that the pathways along the $q\simeq 0.26$ AU are similar because the particles are taken from a narrow region inside the resonance.
We repeat therefore the same integration  (as described in  Section \ref{method}) for other data sets, where the chaotic particles
were selected  for different $(i, \Omega, \omega, M)$ combinations. It should be noted that the peculiar structures visible on Fig.\ref{figure1}
were not observed in those cases (see Figures provided as supplementary material). We do not go into the  details of statistical properties of those orbits, 
but focus on their final positions in the $(a,e)$ plane. Those positions are distributed in the same fashion as in Fig.\ref{figure6}, including the 
elimination course along  $q\simeq 0.26$ AU\footnote{See bottom panels in Fig. 10 and Fig. 11 in the supplementary material.}.

Thus, our results clearly suggest that the $q\simeq 0.26$ AU line may be a natural inner stability border of the NEO region.
Let us recall a recent work of \cite{Granvik2016}, where it was found that NEOs do not survive certain low values of $q$.
As claimed by the authors, most such removals are credited to super-catastrophic breakups depending largely on the  physical properties of NEOs 
(diameter, masses, albedo). Here we have observed a clear limitation in $q$, although we do not consider any of those physical parameters.
This leads to the conclusion that the disappearance of NEOs at low $q$ may also have a dynamical character. 

According to \cite{Bottke.et.al.2002}, a typical pathway of a test particle starting in the Main Belt could be illustrated by the following scheme:
$$Main Belt \rightarrow Resonances\rightarrow NEO \rightarrow sink $$
where the so-called {\it sink} is a dynamical route along which  most particles escape from the Solar System or fall on the Sun.

If we look at the orbital pathways of our individual test particles reaching low perihelion distances, 
we notice that most of the bodies travel along a direction close to the mentioned $q\simeq 0.26$ AU line. 
Therefore, we identify this line with the so called {\it sink} and modify the above scheme in the following way:
$$Main Belt \rightarrow 5:2 \ resonance \rightarrow NEO \rightarrow q\simeq 0.26 AU. $$

\begin{figure}
 \centering
\includegraphics[height=7.cm, angle =-90]{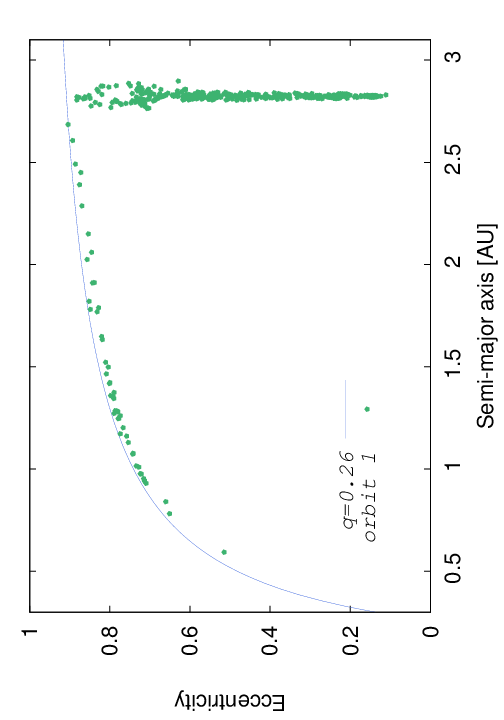}
\includegraphics[height=7.cm, angle =-90]{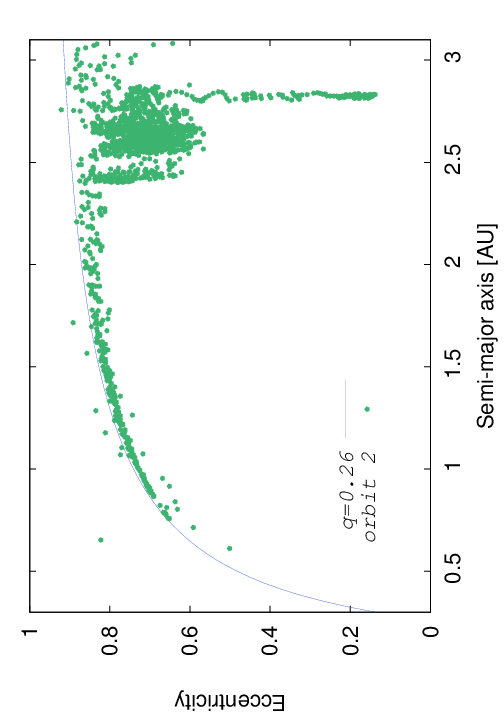}

\includegraphics[height=7.cm, angle =-90]{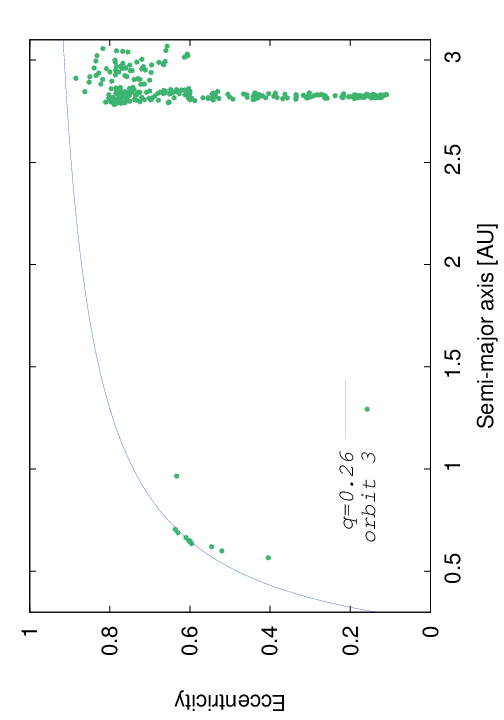}
\includegraphics[height=7.cm, angle =-90]{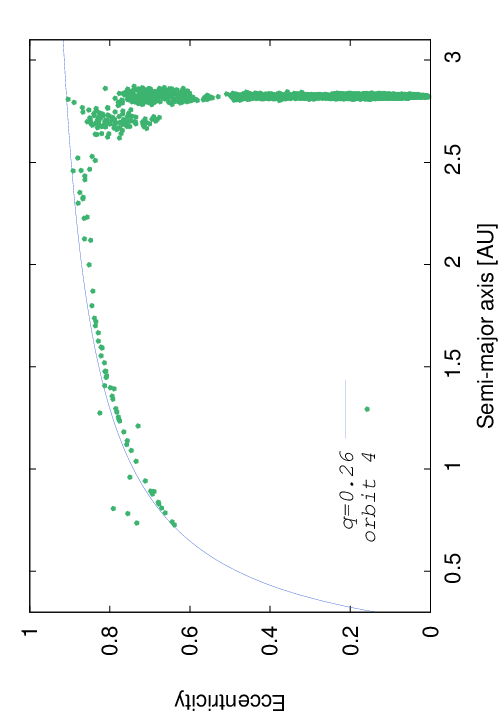}
\caption{Orbital pathways in the $(a,e)$ plane for four different orbits that during some time navigate along the $q\simeq 0.26$ AU course
(full blue line on the above panels). We identify this line with the so-called {\it sink}, a dynamical route that particles follow 
before they are eliminated from the system. The particles on the above figures are randomly chosen among the ones that have reached 
low perihelion distances.}
\label{figure7}
\end{figure}

In order to illustrate this, we have randomly chosen four particles that   reach  low $q$ and have shown their orbital pathways
in the $(a,e)$ plane in Fig.~\ref{figure7}. The $q=0.26 AU$  boundary is marked with a full line on all the four panels. 
Each of the four particles, after reaching high eccentricities and leaving the 5:2 resonance (or the neighboring resonances), 
navigates along a direction close to the $q=0.26 AU$ course and leaves the system. Thus, the direction which we identified as a natural stability border,
at the same time represents a dynamical highway along  which  particles travel to their final destination. 

\subsection{3200 Phaethon}

In this final part, we will count the test bodies recovering the orbit of the asteroid  (3200) Phaethon.
That is, we search in the $(a,e,i)$ element space those particles that at some moment of the integration 
satisfy the following criteria  $|a-a_{Ph}|< 0.1 $,  $|e-{e_Ph}|< 0.1 $ and  $|i-i_{Ph}|< 3$, where 
 $(a_{Ph}, e_{Ph}, i_{Ph})$ are the semi-major axis, eccentricity and inclination of the asteroid (3200) Phaethon
whose values are  $(a_{Ph}, e_{Ph}, i_{Ph})=(1.271,\ 0.889,\ 22.243)$.
 
According to the results in \cite{Bottke.et.al.2002} Phaethon had a 0 probability of coming from the
Jupiter family comets or Outer Belt region. This probability increased to 1\% in the work of \cite{deLeon2010}.
Among our 1000 test particles, we have found 80 bodies satisfying the above criteria, increasing this probability 
up to 8\%. 

\section{Conclusions}
\label{concl}

The main results of this paper are summarized below. 

\begin{itemize}

\item We have computed dynamical maps of the 5:2 MMR for Solar System models with and without the inner planets. 
A direct comparison between them  enabled a detection of weak neighboring resonances with  inner planets.

\item We have  observed chaotic structures inside the 5:2 resonance and we have illustrated that those 
structures represent a very effective source of transportation processes. 

\item Time scales of the 5:2 MMR removal abilities  are not significantly shortened, but the amount of material  
becoming NEOs (99.5\% of the test bodies), reaching semi-major axis below 1 AU (23.4\%) or entering the Hill sphere of Earth (57.4\%)
show a large disagreement with earlier studies. However, we point out that this mismatch is primarily attributed to the choice of
initial conditions which are selected  intentionally along the most unstable parts of the resonance that should have 
highly efficient transportation abilities.

\item The final destinations in the $(a,e)$ plane suggest that  in addition to the main removal course caused by Jupiter, 
most particles are ejected out from an unknown direction defined with $q\simeq 0.26$ AU. Dynamical origin of this 
elimination line requires further investigation.

\item  The percentage of our test particles that recovered a Phaethon like orbit is 8\%. 

\end{itemize}

Using sensitive numerical methods, we have shown that the 5:2 MMR with Jupiter has strong 
dynamical removal abilities that  can drive a large amount of bodies in the close neighborhood of Earth. 
This result could explain the growing number of  NEOs and meteorites that are identified as former members 
of the asteroid families located in the Outer Belt.

Applying  this  method on a single resonance enabled us to observe the 'hidden potential' of 
its transportation abilities.  Although the above results have to be analyzed  further in future work, 
we conclude that extending the  same method to other  regions in the Solar System  and  other resonances
could  provide a  clearer  insight and a better understanding of the orbital migration phenomena. 

\section*{Acknowledgements}
This research was supported by the Ministry of Education, Science and Technological Development 
of the Republic of Serbia, under the project 176011 'Dynamics and kinematics of celestial 
bodies and systems'. The calculations were performed on a Fermi cluster located 
at the Astronomical Observatory of Belgrade, purchased by the project
III44002 'Astroinformatics: Application of IT in astronomy and close fields'.
The author is very grateful to the editor and an anonymous reviewer for their careful reading 
and constructive comments, which helped to improve the manuscript.

\onecolumn

\twocolumn






\label{lastpage}
\end{document}